\documentclass{ws-mpla}

\begin{document}

\markboth{S. I. Bastrukov,
  H.-K. Chang,  E.-H. Wu and I. V. Molodtsova }
{Self-gravitating astrophysical mass with singular central density}

%%%%%%%%%%%%%%%%%%%%% Publisher's Area please ignore %%%%%%%%%%%%%%
\catchline{}{}{}{}{}
%%%%%%%%%%%%%%%%%%%%%%%%%%%%%%%%%%%%%%%%%%%%%%%%%%%%%%%%%%%%%%%%%%%

\title{Self-gravitating astrophysical mass with singular central density vibrating in fundamental mode}

\author{\footnotesize S.I. BASTRUKOV\footnote{Joint Institute for
 Nuclear Research,
141980 Dubna, Russia},\, H.-K. CHANG,\, E.-H. WU}

\address{
Department of Physics and  Institute of Astronomy,
  National Tsing Hua University, Hsinchu, 30013, Taiwan\\
 bast@phys.nthu.edu.tw; hkchang@phys.nthu.edu.tw}

\author{I. V. MOLODTSOVA}

\address{Joint Institute for
 Nuclear Research,
141980 Dubna, Russia\\
molod@theor.jinr.ru
}

\maketitle

\pub{Received (Day Month Year)}{Revised (Day Month Year)}

\begin{abstract}
   The fluid-dynamical model of a self-gravitating mass of viscous liquid with singular density at the center vibrating in fundamental mode is considered
   in juxtaposition with that for Kelvin fundamental mode in a homogeneous heavy mass of incompressible inviscid
   liquid. Particular attention is given to the difference between spectral
   formulae for the frequency and lifetime of $f$-mode in the singular and homogeneous models.
   The newly obtained results are discussed in the context of theoretical asteroseismology of pre-white dwarf stage of red giants and stellar cocoons -- spherical gas-dust clouds with dense star-forming core at the center.

\end{abstract}

\ccode{PACS 04.40.-b.}

%\vspace{0.5cm}

\section{Introduction}
  Studying highly idealized models of vibrating self-gravitating fluid masses forms one of important issues of theoretical asteroseismology which seeks to explain variability in brightness of an observable object as being produced by its vibrations. Among central to this domain of astrophysics is Kelvin fluid-mechanical model of a heavy mass of inviscid incompressible liquid of uniform density $\rho$ undergoing free oscillations with nodeless irrotational velocity of fluctuating flow\cite{K-1863}. An outstanding importance of this fiducial model, whose extended discussion can be found in Refs. [2-7], is that it
  sets the standard for analytic study of non-radial pulsations of the main-sequence stars and serves as an example, according to Chandrasekhar\cite{CH-61}, of "at least one problem for which analytic solution can be found" and which illustrates "the type of difficulties one must confront in the other problem", because "in most instances, the problems become of such complexity and involve so many parameters that elementary methods of solution seem impracticable". It is this model from which the very notion of fundamental vibrational
  mode has come into the theory of stellar pulsations\cite{AG-97}.

  The characteristic peculiarity of a liquid star vibrating in fundamental mode is that
  the velocity of oscillating flow is described by the potential vector field whose potential obey the Laplace equation of the form
  \begin{eqnarray}
  \label{e1.2}
  && \nabla\cdot  \delta {\bf v}({\bf r},t)=0,\quad
  \delta {\bf v}({\bf r},t)=\nabla \chi({\bf r},t),\\
  \label{e1.3}
  && \nabla^2 \chi({\bf r},t)=0,\quad
  \chi({\bf r},t)=A_\ell\,r^\ell\,P_\ell(\cos\theta){\dot \alpha}(t).
  \end{eqnarray}
  The last equation exhibits node-free character of the velocity field as a function of distance from center to the surface of the star. It is this feature of oscillating flow is regarded as the major kinematic
  signature of fundamental vibration mode. In the Kelvin model of homogeneous mass of a heavy incompressible liquid, implying that the density is not altered, $\delta \rho=0$, the above equations emerge from the continuity equation
  \begin{eqnarray}
 \label{e1.1}
 \delta{\dot \rho}=-\nabla \,(\rho\,\delta {\bf v})=-(\delta {\bf v}\cdot\nabla)\,\rho-\rho\,
 \nabla\cdot \delta {\bf v}.
 \end{eqnarray}
 Our purpose here is to explore some peculiarities of the fundamental vibration mode in astrophysical object
 with inhomogeneous distribution of mass.
 In so doing  we consider a model of highly inhomogeneous stellar object in which the non-uniform density profile has singularity at the star center of the form
\begin{eqnarray}
 \label{e1.4}
 \rho(r)=\frac{5}{6}\rho_s \left({\frac{R}{r}}\right)^{1/2}.
 \end{eqnarray}
 One of the most conspicuous features of this density profile
 is that the total mass of such a star
 \begin{eqnarray}
 \label{e1.5}
 M=\int \rho(r)\,d{\cal V}= \frac{4\pi}{3}\rho_s R^3
 \end{eqnarray}
 is finite and identical to that for the total mass of the canonical homogenous liquid star model.
 In somewhat different context the model of inhomogeneous liquid star with similar density profile has been briefly discussed by Clayton\cite{CL-86}.
 This curious feature of the model, which from now on is referred to as the singular star model,
 is interesting in its own right because it permits too analytically tractable solution of the eigenfrequency problem for $f$-mode. Perhaps the most striking feature distinguishing
 vibrational behavior of inhomogeneous from homogeneous mass is that the nodeless non-rotational vibrations of a spherical liquid mass of non-uniform density are of substantially compressional character, as it follows from the continuity equation. One of the prime purposes of our study here, therefore,
 is to find out how this striking distinction between the density profiles is reflected in the frequency and lifetime spectra of fundamental vibration mode.

 The paper is organized as follows. In section 2, general fluid-mechanical equations of nodeless stellar vibrations in fundamental mode are outlined.
 In section 3, the detailed analytic computation of the spectral equations for the frequency and lifetime of $f$-mode in the singular liquid star model is presented followed by their comparison with those for the Kelvin fundamental mode in the canonical homogeneous liquid star model which is the subject of Section 4.
 The newly obtained results are highlighted in section 5 and
 briefly discussed in the context of asteroseismology of pre-white dwarf stage of red giants and stellar cocoons -- spherical gas-dust clouds with dense star-forming core at the center.

\section{General equations of nodeless stellar vibrations in fundamental mode}

 The state of motion of flowing stellar matter under the action of forces of
 buoyancy, gravity and viscous stress is uniquely described in terms of five dynamical variables, to wit, the density $\rho({\bf r},t)$, three components
 of the flow velocity ${\bf v}({\bf r},t)$ and the pressure $p({\bf r},t)$ obeying the coupled equations of fluid-mechanics and Newtonian universal gravity
 \begin{eqnarray}
 \label{e2.1}
 &&\frac{d\rho}{dt}=-\rho\,
 \frac{\partial v_k}{\partial x_k},\quad\quad \frac{d}{dt}=
 \frac{\partial }{\partial t}+v_k\frac{\partial }{\partial x_k},
 \\[0.2cm]
 \label{e2.2}
 && \rho\frac{dv_i}{dt}=-
 \frac{\partial p}{\partial x_i}+\rho\frac{\partial U}{\partial x_i}+
 \frac{\partial \pi_{ik}}{\partial x_k},
 \\[0.2cm]
 \label{e2.3}
 && \frac{dp}{dt}=\frac{\Gamma_1 p}{\rho}\frac{d\rho}{dt}=-\Gamma_1\,p\frac{\partial v_k}{\partial x_k}, \quad \quad \Gamma_1=\frac{\partial \ln p}{\partial \ln \rho},
  \\[0.2cm]
 \label{e2.4}
 &&\nabla^2 U=-4\pi G\rho.
 \end{eqnarray}
 The equation for density $\rho$ is the condition of continuity of flowing stellar matter. The Navier-Stokes equation for the velocity flow expresses the second law of  Newtonian dynamics
 for viscous liquid and equation for pressure is the condition of adiabatic behavior of stellar matter; $\Gamma_1$ stands for the adiabatic coefficient of gaseous pressure in the star. This latter
 condition means that in the process of motion, like vibrations, the time scale of energy exchange between infinitesimally close elementary volume of liquid is much longer than characteristic period of oscillations, see, for example, Refs. [11,12]).
 The tensor of Newtonian viscous stresses is given by
 \begin{eqnarray}
  \label{e2.5}
  \pi_{ik}=2\eta\,v_{ik}+\left(\zeta-\frac{2}{3}\,\eta\right)v_{jj}\,\delta_{ik},\quad\quad
  v_{ik}=\frac{1}{2}\left(\nabla_k v_i+\nabla_i v_k\right).
  \end{eqnarray}
  In computing frequency of fundamental vibration mode, the equilibrium density profile
  $\rho$, the pressure $P_c$ at the star center as well as the transport coefficients of stellar
  matter, the shear $\eta$ and the bulk $\zeta$ viscosities, are regarded
  as input, in advance given, parameters.

  The potential of self-gravity $U(r)$ and pressure $p(r)$ in motionless, ${\bf v}=0$,
  state of hydrostatic equilibrium, are the solutions of coupled equations
  \begin{eqnarray}
 \label{e2.6}
 \nabla_i\,p(r)=\rho(r)\nabla_i\, U(r),\quad\quad \nabla^2 U(r)=-4\pi G\rho(r).
 \end{eqnarray}
 The gravity potential $U$ is determined, in effect, by Poisson equation for internal
 potential $U_i$ and Laplace equation for external one $U_e$
 supplemented by the standard boundary conditions of the continuity of these potentials and their normal
 derivatives on the star surface
 \begin{eqnarray}
 \label{e2.7}
 && \nabla^2 U_i(r)= - 4\pi G \rho(r),\quad r\leq R,\quad
 \quad\nabla^2 U_e(r)=0,\quad r>R,\\[0.2cm]
\label{e2.8}
 && U_i(r)=U_e(r)\Big\vert_{r=R},\quad\quad
 \frac{dU_i(r)}{dr}=\frac{dU_e(r)}{dr}\Big\vert_{r=R}.
 \end{eqnarray}
 The general solution of equation for pressure is specified by standard boundary condition of stress-free surface $p(r=R)=0$.

 The equations of linear oscillations of stellar matter about
 stationary state of hydrostatic equilibrium of a star with non-uniform equilibrium density, $\rho=\rho(r)$, are obtained by applying to (\ref{e2.1})-(\ref{e2.4}) the standard procedure of linearization
 \begin{eqnarray}
 \label{e2.9}
 && \rho\to\rho(r)+\delta \rho({\bf r},t),\quad  p\to p(r)+\delta p({\bf r},t),\\
 \label{e2.10}
 && v_i\to v_i+\delta v_i({\bf r},t),\quad [v_i=0],\quad U\to U ({\bf r})+\delta U({\bf r},t).
 \end{eqnarray}
 As was stated, in this work we focus on the regime of irrotational vibrations in which the velocity field
 of fluctuating flow subjects to
 \begin{eqnarray}
 \label{e2.11}
 \nabla\cdot \delta {\bf v}=0,\quad  \nabla\times \delta {\bf v}=0.
 \end{eqnarray}
 Given this, the continuity equation takes form
 \begin{eqnarray}
 \label{e2.12}
 \delta{\dot \rho}=-(\delta v_k\nabla_k)\rho.
 \end{eqnarray}
 The liner fluctuations of the flow velocity are governed by linearized Navier-Stokes equation
 \begin{eqnarray}
 \label{e2.13}
 &&\rho\delta {\dot v}_i=-
 \nabla_i\delta p+\rho\nabla_i\delta U+\delta\rho\nabla_i U
 +\nabla_k \delta\pi_{ik},
 \\[0.2cm]
 \label{e2.14}
 &&\delta \pi_{ik}=2\eta \delta v_{ik},\quad\quad
 \delta v_{ik}=\frac{1}{2}(\nabla_i\delta v_k+\nabla_k\delta v_i).
 \end{eqnarray}
 From (\ref{e2.3}) it follows that rate of change in the pressure is controlled by equation
 \begin{eqnarray}
 \label{e2.15}
 \delta {\dot p}=-(\delta v_k\nabla_k)\,p.
 \end{eqnarray}
 Fluctuations in potential of self-gravity inside the star $\delta U=\delta U_i$, caused by fluctuations in density $\delta \rho$, subject to the Poisson equation
  \begin{eqnarray}
 \label{e2.16}
 \nabla^2 \delta U=-4\pi G\,\delta \rho.
 \end{eqnarray}
 The energy balance in the process of oscillations is controlled by equation
 \begin{eqnarray}
 \label{e2.17}
 \frac{\partial }{\partial t}\int\frac{\rho\delta v^2}{2}d{\cal V}=-\int[(\delta v_k\nabla_k)\delta p-\rho(\delta v_k\nabla_k)\delta U-\delta\rho(\delta v_k\nabla_k)U-\delta\pi_{ik}\delta v_{ik}]
  d{\cal V}
 \end{eqnarray}
 which is obtained after scalar multiplication of (\ref{e2.13}) by $\delta v_i$
 and integration over the star volume.
 To compute the eigenfrequency of vibrations we take advantage of the Rayleigh's energy variational principle at the base of which lies the following separable representation of fluctuating
 variables
 \begin{eqnarray}
 \label{e2.18}
 && \delta v_i({\bf r},t)=a_i({\bf r}){\dot \alpha}(t),\quad \delta U({\bf r},t)=\phi({\bf r}){\alpha}(t),\\
 \label{e2.19}
 && \delta \rho({\bf r},t)=\tilde \rho({\bf r})\,{\alpha}(t),\quad \tilde \rho({\bf r})=-(a_k({\bf r})\nabla_k)\,\rho(r),\\
 \label{e2.20}
 &&\delta p({\bf r},t)=\tilde p({\bf r})\,{\alpha}(t),\quad \tilde p({\bf r})=-(a_k({\bf r})\nabla_k)\,p(r),\\
 \label{e2.21}
 && \delta \pi_{ik}({\bf r},t)=2\eta\,\delta v_{ik}=2\eta\,a_{ik}({\bf r}){\dot \alpha}(t),\\
 \label{e2.22}
 && \delta v_{ik}({\bf r},t)=a_{ik}({\bf r}){\dot \alpha}(t),
 \quad\quad a_{ik}({\bf r})=\frac{1}{2}(\nabla_ia_k({\bf r}),
 +\nabla_k a_i({\bf r})).
 \end{eqnarray}
 Hereafter $a_i({\bf r})$ stands for the time-independent field of instantaneous material
 displacements and $\alpha(t)$ for the temporal amplitude of oscillations.
 The key idea of such representation is that it transforms equation of energy
 balance (\ref{e2.17}) into equation for $\alpha$  having well-familiar form
 of equation of damped oscillations
  \begin{eqnarray}
 \label{e2.23}
 && \frac{d{\cal E}}{dt}=-2{\cal F},\quad\quad
 {\cal E}=\frac{{\cal M}{\dot\alpha}^2}{2}+\frac{{\cal K}\alpha^2}{2},\quad
 {\cal F}=\frac{{\cal D}{\dot \alpha^2}}{2},
 \\[0.2cm]
 \label{e2.24}
 && {\cal M}{\ddot \alpha}+{\cal D}{\dot \alpha}+{\cal K}\alpha=0,
 \\[0.2cm]
 \label{e2.25}
 && {\cal M}=\int \rho(r) a_k \,a_k\,d{\cal V},\quad {\cal D}=2\int \eta(r) a_{ik}\,a_{ik}\,d{\cal V},\\
 && {\cal K}=\int[(a_k\nabla_k\rho(r))(a_k\,\nabla_k U({\bf r}))-\rho(r)(a_k\nabla_k)\phi({\bf r})-(a_k\nabla_k)(a_k\nabla_k)p(r)]d{\cal V}.
 \end{eqnarray}
 Here ${\cal E}$ is the energy of free, non-dissipative, oscillations and ${\cal F}$ is the dissipative
 function of Rayleigh describing their damping by shear viscosity of stellar matter.
 The solution of (\ref{e2.24}) is given by
 \begin{eqnarray}
 \label{e2.26}
 && \alpha(t)=\alpha_0\,\exp(-t/\tau)\,\cos \omega(\tau) t,
 \\[0.2cm]
 \label{e2.27}
 && \omega^2(\tau)=\omega^2[1-(\omega\tau)^{-2}],
 \quad\omega^2=\frac{{\cal K}}{{\cal M}},\quad
  \tau=\frac{2{\cal M}}{{\cal D}}
 \end{eqnarray}
  where $\omega(\tau)$ is the frequency of dissipative oscillations damped by viscosity, $\omega$ is the
 frequency of free oscillations and $\tau$ is their lifetime. Thus, to compute
 the frequency and lifetime one need to specify all variables entering integral parameters of inertia
 ${\cal M}$, stiffness ${\cal K}$ and viscous friction ${\cal D}$.

 We start
 with the field of instantaneous displacements $a_k({\bf r})$ which is
 the key kinematic characteristics of $f$-mode.
 In the star undergoing irrotational oscillations in the $f$-mode, the shape of an arbitrary spherical surface takes the form of harmonic spheroids which are described by
 \begin{eqnarray}
 \label{e2.28}
 r(t) = r [ 1+ \alpha_\ell (t) P_\ell (\zeta)],\quad \zeta=\cos\theta
\end{eqnarray}
 where $P_\ell (\zeta)$ is the Legendre polynomial of the multipole order $\ell$ specifying
 the overtone number in fundamental vibration mode.
 In the system with fixed polar axis the potential field of velocity is found from Laplace equation
 supplemented by boundary condition that radial component of velocity on the star surface equals
 to the rate of the surface distortions taking the shape of harmonic spheroids
 \begin{eqnarray}
 \label{e2.29}
 && \nabla_k\,\delta v_k=0,\quad \delta v_k=\nabla_k \chi,\quad \nabla^2 \chi=0\quad\to\quad \chi=A_\ell\, r^\ell\, P_\ell(\zeta)\,{\dot\alpha}(t),\\
 \label{e2.30}
 && \delta v_r\big\vert_{r=R}={\dot R}(t),\quad  R(t) = R [ 1+ \alpha_\ell(t) P_\ell (\zeta)].
\end{eqnarray}
Taking into account that $\delta {\bf v}({\bf r},t)={\bf a}({\bf r}){\dot\alpha}(t)$ one has
 \begin{eqnarray}
 \label{e2.31}
 {\bf a}({\bf r})=A_\ell\nabla r^\ell\,P_\ell(\zeta),\quad\quad
 A_\ell=\frac{1}{\ell R^{\ell-2}}.
\end{eqnarray}
In terms of this field, the linearized equations for the density $\delta \rho$ and for the
pressure $\delta p$ reads
\begin{eqnarray}
 \label{e2.32}
 && \delta \rho({\bf r},t)=\tilde \rho({\bf r})\,{\alpha}(t),\quad  \tilde \rho({\bf r})=-({\bf a}
 \cdot \nabla)\,\rho(r)=-A_\ell P_\ell(\zeta) \ell\,r^{\ell-1}\frac{\partial \rho(r)}{\partial r},\\
 \label{e2.33}
 && \delta p({\bf r},t)=\tilde p({\bf r})\,{\alpha}(t),\quad  \tilde p({\bf r})
 =-({\bf a}\cdot \nabla)\,p(r)=-A_\ell P_\ell(\zeta) \ell\,r^{\ell-1}\frac{\partial p(r)}{\partial r}
 \end{eqnarray}
where $\rho(r)$ and $p(r)$ are the density and pressure of gravitationally equilibrium, hydrostatic, configuration.

 To compute variations in the potential of gravity one must consider
 two equations for internal $\delta U_i$ and external $\delta U_i$ potentials
 \begin{eqnarray}
 \label{e2.34}
 && \nabla^2 \delta U_i({\bf r},t)=-4\pi\,G\delta \rho({\bf r},t),\quad  \nabla^2 \phi_i({\bf r})=-4\pi\,G \tilde \rho({\bf r}),\\
  \label{e2.35}
 && \nabla^2 \delta U_e({\bf r},t)=0,\quad  \nabla^2 \phi_e({\bf r})=0.
 \end{eqnarray}
The outlined energy variational method provides a general framework for computing
the frequency of $f$-mode in a inhomogeneous Newtonian liquid star with arbitrary form of
non-uniform density profile.

\section{Fundamental vibration mode in stellar objects with singular density at the center}

 In what follows we use term singular stellar object or, for brevity, singular star for a self-gravitating mass of viscous
 liquid with the non-uniform and singular in the star center density profile given
 by equation (\ref{e1.4}) whose total mass is identical to that for homogenous star model.
 For such a singular star, the equilibrium potentials and fields of universal gravity are given by
 \begin{eqnarray}
  \label{e3.1}
 && \nabla^2 U_i(r)=-4\pi G \rho(r),\quad \rho(r)=\frac{5}{6}\rho_s \sqrt{\left({\frac{R}{r}}\right)},
 \,M=\int \rho(r)\,d{\cal V}= \frac{4\pi}{3}\rho_s R^3,\\
 \label{e3.2}
 && U_i(r<R)=\frac{20\pi}{9}G\rho_s R^2\left[1-\frac{2}{5}
 \left(\frac{r}{R}\right)^{3/2}\right]=\frac{5}{3}\frac{G M}{R}\left[1-\frac{2}{5}
 \left(\frac{r}{R}\right)^{3/2}\right],\\
 \nonumber
 &&{\rm \bf g}_i(r<R)=-\nabla U_i=
 \frac{G M}{R^3}\left(\frac{r}{R}\right)^{1/2}{\bf r},\\[0.2cm]
 \label{e3.3}
 && U_e(r>R)=\frac{4\pi}{3}G\rho_s \frac{R^3}{r}=\frac{G M}{r},\,\, {\rm \bf g}_e(r>R)=-\nabla U_e=\frac{G M}{r^3} {\bf r}.
 \end{eqnarray}
 The hydrostatic pressure obeying the boundary condition of free-from-stress surface, $p(r)\vert_{r=R}=0$, reads
 \begin{eqnarray}
 \label{e3.4}
 p(r)=\frac{10\pi}{9}G\rho_s^2R\left(R-r\right)=P_c\left(1-\frac{r}{R}\right),\quad P_c=\frac{10\pi}{9}G\rho_s^2R^2=\frac{5}{8\pi}\frac{G M^2}{R^4}.
 \end{eqnarray}
 It is remarkable that the pressure at the star center, where the density has singularity,
 is finite and its radial profile is the linear function of distance from the center to the star surface. The pressure in the center is defined by equation of state of the stellar matter\cite{PHIL-94}. With in advance given equation of state for $P_c$ the rightmost identity
 in (\ref{e3.4}) is considered as definition of the star radius. The internal gravitational energy is
 \begin{eqnarray}
 \label{e3.5}
 W_s=\frac{1}{2}\int \rho_s\, U_i d{\cal V}=\frac{11}{18}\frac{GM^2}{R}\simeq 1.02 W,\quad W=\frac{3}{5}\frac{GM^2}{R}
  \end{eqnarray}
where $W$ is the total gravitational energy of homogeneous star of equivalent mass $M$.
To take into account the compression effect of self-gravity on mechanical property of star matter we adopt that radial profile of viscosity coefficient is identical to that for the equilibrium pressure profile, that is, of the form
\begin{eqnarray}
 \label{e3.6}
 \nu(r)=\nu_c\left(1-\frac{r}{R}\right)
 \end{eqnarray}
where $\nu_c$ is the shear viscosity in the star center which along with $\rho_s$ and
$P_c$ are regarded as input parameters of the model.

\subsection{Exact solution of Poisson equation for variations of self-gravity potential in singular star}

 Having defined the equilibrium profiles of density $\rho(r)$, the pressure $p(r)$, the shear viscosity profile $\nu(r)$ and knowing the field of instantaneous displacements
 ${\bf a}({\bf r})$ we are able to compute the inertia ${\cal M}$ and viscous friction ${\cal D}$.
 However, in order to compute the stiffness ${\cal K}$ we must calculate fluctuations in the potential
 of self-gravity $\delta U({\bf r},t)=\phi({\bf r})\alpha(t)$, that is, to solve Poisson equation
 for $\phi({\bf r})$ with a fairly non-trivial right part
  \begin{eqnarray}
 \label{e3.7}
 &&\nabla^2 \phi_i(r,\zeta)=-\frac{5\pi}{3} A_{\ell}\, G \rho_s\, R^{1/2}\,\ell\, r^{\ell-5/2} P_{\ell}(\zeta).
 \end{eqnarray}
 In the spherical polar coordinates we have
 \begin{eqnarray}
 \label{e3.8}
 \frac{1}{r^2}\frac{\partial }{\partial r}r^2\frac{\partial \phi_i(r,\theta)}{\partial r}+\frac{1}{r^2\sin\theta}\frac{\partial }{\partial \theta}\sin\theta\frac{\partial \phi_i(r,\theta)}{\partial \theta}&=&\\ \nonumber
        &-&\frac{5\pi}{3}\,A_{\ell}\, G \rho_s\, R^{1/2}\,\ell\, r^{\ell-5/2} P_{\ell}(\cos\theta).
 \end{eqnarray}
 Assuming a solution of the form
 \begin{eqnarray}
 \label{e3.9}
 && \phi_i(r,\theta)=\frac{u(r)}{r}\,P_\ell(\cos\theta)
 \end{eqnarray}
 and taking into account that $P_\ell(\cos\theta)$ is the solution of equation
 \begin{eqnarray}
 \label{e3.10}
 && \frac{1}{\sin\theta}\frac{\partial }{\partial \theta}\sin\theta\frac{\partial P_\ell(\cos\theta)}{\partial \theta}=-\ell(\ell+1)\,P_\ell(\cos\theta)
 \end{eqnarray}
 we obtain
 \begin{eqnarray}
 \label{e3.11}
 && r^2\frac{\partial^2 u(r)}{\partial r^2}-\ell(\ell+1)u(r)=-\frac{5\pi}{3}\,A_\ell\, G \rho_s\, R^{1/2}\,\ell\, r^{\ell+1/2}.
 \end{eqnarray}
 The general solution of equation for $\phi_i$, which is finite at the origin, is given by
\begin{eqnarray}
\label{e3.12}
&& \phi_i(r,\theta)=\left\{C\,r^\ell+\frac{20\pi}{3}\frac{\ell}
{4\ell+1}\,\,A_\ell\,\rho_s\, G\,R^{1/2}r^{\ell-1/2}\right\}P_\ell(\cos\theta)
 \end{eqnarray}
 Outside the star we have
  \begin{eqnarray}
  \label{e3.13}
&&\nabla^2 \delta U_e=0\quad\to\quad \nabla^2 \phi_e({\bf r})=0,\\
\label{e3.14}
&& \phi_e=D\, r^{-\ell -1}P_\ell(\cos\theta).
 \end{eqnarray}
 The arbitrary constants $C$ and $D$ are eliminated from boundary conditions
 \begin{eqnarray}
 \label{e3.15}
\phi_i=\phi_e\Big\vert_{r=R},\quad\quad \frac{\partial \phi_i}{\partial r}=\frac{\partial \phi_e}{\partial r}\Big\vert_{r=R}.
\end{eqnarray}
which yield
 \begin{eqnarray}
 \label{e3.16}
 C=-\frac{10\pi}{3}\,A_\ell\,\rho_s\,G\,\frac{\ell}{2\ell+1},\quad
 D=\frac{10\pi}{3}\,A_\ell\, \rho_s\, G\, \frac{\ell}{(4\ell+1)(2\ell+1)}R^{2\ell+1}.
 \end{eqnarray}
Finally, we obtain
\begin{eqnarray}
\label{e3.17}
&& \phi_i(r,\theta)=-\frac{10\pi}{3}\frac{\ell}{2\ell+1}A_\ell\, \rho_s\, G\,\left[r^\ell
-\frac{2(2\ell+1)}{4\ell+1}R^{1/2}
\,r^{\ell-1/2}\right]P_\ell(\cos\theta).
 \end{eqnarray}
It is worth emphasizing that following this line of argument one can get the solutions
of Poisson equation for a more wide class of inhomogeneous star models undergoing node-free spheroidal vibrations with non-rotational field of velocity.

\subsection{Spectral equations for frequency and lifetime of $f$-mode}

The mass parameter ${\cal M}$ is given by
\begin{eqnarray}
\label{e3.18}
{\cal M}&=&\int_V \rho(r)\,[a^2_r+a^2_\theta]\,d{\cal V}
 =\frac{5\pi}{3}\rho_s A_\ell^2 R^{1/2}\int\limits_{0}^{R}\,r^{2\ell-1/2}\,dr\,\\ \nonumber
  &\times& \int\limits_{-1}^{+1}
\left[\ell^2 P^2_\ell(\zeta)+\left(1-\zeta^2\right)\left(\frac{dP_{\ell}(\zeta)}{d\zeta}\right)^2\right] d\zeta
=\frac{20\pi}{3}\,A_\ell^2\,\rho_s\,R^{2\ell +1}\,\frac{\ell}{4\ell +1}.
\end{eqnarray}
Computation of the viscous friction parameter, with
the non-uniform profile of shear viscosity (\ref{e3.6}), yields
\begin{eqnarray}
\nonumber
 {\cal D}&=& 2 \int\eta(r)\left( a_{rr}^2+a_{\theta\theta}^2+a_{\phi\phi}^2+2 a_{r\theta}^2\right) d{\cal V}
 =8\pi A_\ell^2\int\limits_{0}^{R}  \eta(r) r^{2\ell-2} dr \\ [0.2cm]\nonumber
 &\times& \int\limits_{-1}^{1}\left[
 \ell^2 (\ell^2 -\ell+1) P_\ell(\zeta)^2
 -\ell (\ell+1) \zeta P_\ell(\zeta)\frac{dP_\ell(\zeta)}{d\zeta}+
 \zeta^2\left(\frac{dP_\ell(\zeta)}{d\zeta}\right)^{2}\right. \\ \nonumber
 &+&  \left.
 (\ell-1)^2(1-\zeta^2)\left(\frac{d P_\ell(\zeta)}{d\zeta}\right)^{2}\right]
 d\zeta =8\pi A_\ell^2\,\ell(\ell-1)(2\ell-1)\,\int\limits_{0}^{R}
 \eta(r)\, r^{2\ell-2} dr\\
 \label{e3.19}
 &=& 4\pi
 A_\ell^2\,\eta_c\, R^{2\ell-1}(\ell-1).
\end{eqnarray}
The lifetime, $\tau=2{\cal M}/{\cal D}$, of $\ell$-pole overtone of $f$-mode is given by
\begin{eqnarray}
 \label{e3.20}
 && \tau_f(\ell)=\frac{10}{3}\tau_\nu \frac{\ell}{(4\ell+1)(\ell-1)},\quad \tau_\nu=\frac{R^2}{\nu},\quad\nu=\frac{\eta_c}{\rho_s}.
 \end{eqnarray}
Somewhat lengthy but simple calculation of integral parameter of stiffness ${\cal K}$
which is presented in Appendix yields
\begin{eqnarray}
\label{e3.25}
{\cal K}=\frac{40\pi^2}{9}A_\ell^2G\rho_s^2R^{2\ell+1}\frac{\ell(4\ell^2+2\ell-1)}{(2\ell+1)(4\ell+1)}.
 \end{eqnarray}
 From analytic form of ${\cal M}$ and ${\cal K}$ it follows that in the singular star model under consideration the lowest overtone of $f$-mode is of dipole degree, $\ell=1$.

 \begin{figure}[h]
 \centering{\includegraphics[width=11cm]{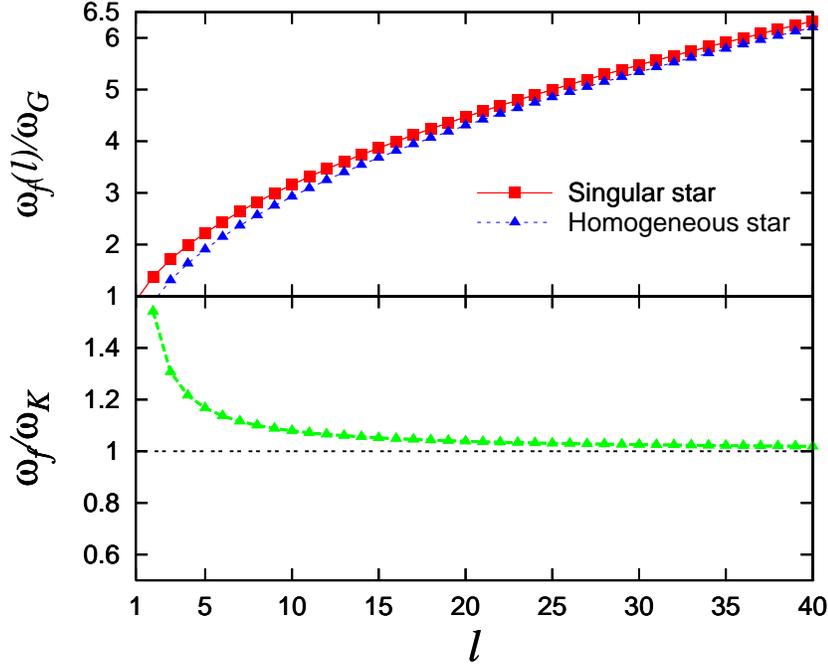}}
 \caption{Frequency of fundamental modes $\omega_f$ as a function of multipole degree $l$ of global node-free  irrotational vibrations of singular and homogeneous star normalized to
 the natural unit of frequency $\omega_G$ (upper panel), and the ratio of spectral equations for the frequency of fundamental mode in singular model $\omega_f$ and in the
 Kelvin homogeneous model $\omega_K$.}
 \end{figure}

 The frequency spectrum of fundamental vibration mode in the singular star model
 reads
\begin{eqnarray}
\label{e3.26}
\omega_f^2(\ell)=\omega^2_G\frac{2\ell(2\ell+1)-1}{2(2\ell+1)},\quad\quad \omega^2_G=\frac{4\pi}{3}\,G\,\rho_s=\frac{GM}{R^3},\quad\quad \ell\geq 1.
 \end{eqnarray}
This last equation can be recast in the following equivalent form
\begin{eqnarray}
\label{e3.27}
&&\omega_f(\ell)=\omega_G\left[\ell-\frac{1}{2(2\ell+1)}\right]^{1/2},\\
&&\ell >> 1 \quad\quad \omega_f(\ell)\simeq\omega_G\sqrt{\ell}.
 \end{eqnarray}
 showing asymptotic behavior of the frequency at very high overtones.

 The obtained frequency spectrum of $f$-mode in singular star model has one and the same physical meaning
 as Kelvin spectral formula for fundamental vibration mode in homogeneous liquid star model
 \begin{eqnarray}
 \label{e3.28}
\omega_K(\ell)=\omega_G\left[\frac{2\ell(\ell-1)}{(2\ell+1)}\right]^{1/2},\quad\quad\omega_G=\sqrt{\frac{GM}{R^3}},\quad \quad\ell\geq 2
 \end{eqnarray}
 but the lowest overtone of Kelvin $f$-mode is of quadrupole degree, $\ell=2$.
 As is demonstrated in Fig.1, the most essential differences between the above spectral equations
 are manifested in low-overtone domain (upper panel) and that at large $\ell$ the frequency spectra of $f$-mode in singular and homogeneous star models shear identical asymptotic behavior (lower panel).
 In the next section this last spectral formula is briefly recovered by the above expounded
 method with allow for the effect of viscous damping of $f$-mode whose lifetime is computed with non-uniform profile of shear viscosity.

\section{Kelvin $f$-mode in the homogeneous liquid star}

 In the canonical homogeneous star model of uniform density, $\rho={\rm constant}$
 the total mass has one and the same form as in above singular model, $M=(4\pi/3)\rho R^3$.
 The internal and external
 potentials of Newtonian gravitational field are the solutions of Poisson equation inside and Laplace equation outside the star
 \begin{eqnarray}
 \label{e4.1}
 && \nabla^2 U_i=-4\pi\,G\rho:\,\, U_i= 2\pi G\rho R^2\left[1-\frac{1}{3}\left(\frac{r}{R}\right)^2\right]=
 \frac{3}{2}\frac{G M}{R}\left[1-\frac{1}{3}\left(\frac{r}{R}\right)^2\right], \\
 \label{e4.2}
 && W=\frac{1}{2}\int \rho\, U_i\, d{\cal V}=\frac{3}{5}\frac{GM^2}{R},\quad \nabla^2 U_e=0:\,\,  U_e(r>R)=\frac{4\pi}{3}G\rho \frac{R^3}{r}=\frac{G M}{r},\\
   \label{e4.3}
 &&{\rm \bf g}=-\nabla U:\quad {\rm \bf g}_i(r<R)=\frac{G M}{R^3}\,{\bf r},\quad
 {\rm \bf g}_e(r>R)=\frac{G\,M}{r^3}\,{\bf r}.
  \end{eqnarray}
 In (\ref{e4.2}), $W$ is the internal gravitational energy $W$ of homogeneous star.
 The solution of equation of hydrostatic for pressure $p(r)$ is given by
 \begin{eqnarray}
 \label{e4.4}
 && \nabla p(r)=\rho\nabla U_i(r)\quad\to\quad
 p(r)=\frac{2\pi}{3}G\rho^2(R^2-r^2)=P_c\left[1-\left(\frac{r}{R}\right)^2\right],\\
 \label{e4.5}
 && P_c=\frac{2\pi}{3}G\rho^2R^2=\frac{3}{8\pi}\frac{G M^2}{R^4},\quad R=\sqrt{\frac{3P_c}{{2\pi}{G\rho^2}}}.
 \end{eqnarray}
 The last expression shows again that the star radius is determined by the central pressure $P_c$ related to the density $\rho$ by equation of state.

 \begin{figure}[h]
 \centering{ \includegraphics[width=6.5cm]{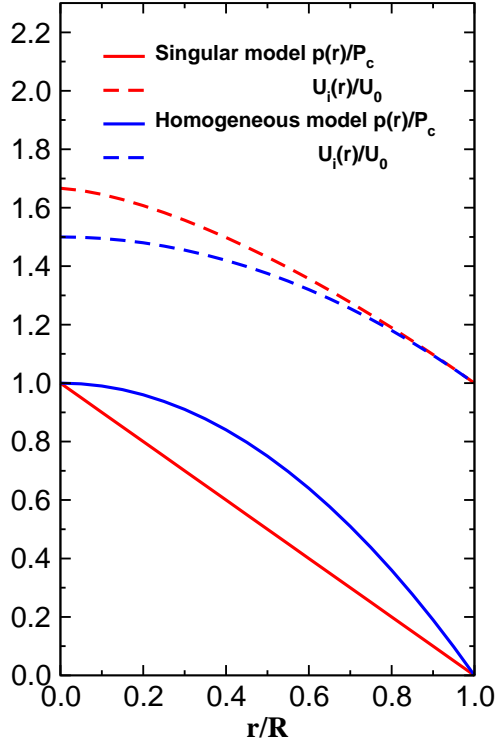}}
 \caption{
 The fractional pressure and gravity potential profiles in singular inhomogeneous star model
 and in the canonical homogenous star model.}
\end{figure}

 For comparison, in Fig.2 we plot the fractional pressure and gravity potential profiles
 computed in homogeneous model
\begin{eqnarray}
\label{e4.6}
 \frac{p(r)}{P_c}=\left[1-\left(\frac{r}{R}\right)^2\right],\,  P_c=\frac{3}{8\pi}\frac{G M^2}{R^4},\quad  \frac{U_i(r)}{U_0}=\frac{3}{2}\left[1-\frac{1}{3}
 \left(\frac{r}{R}\right)^{2}\right],\,U_0=\frac{G M}{R}
 \end{eqnarray}
  and in singular inhomogeneous model
 \begin{eqnarray}
 \label{e4.7}
 \frac{p(r)}{P_c}=\left(1-\frac{r}{R}\right),\quad P_c=\frac{5}{8\pi}\frac{G M^2}{R^4},\,
 \frac{U_i(r)}{U_0}=\frac{5}{3}\left[1-\frac{2}{5}
 \left(\frac{r}{R}\right)^{3/2}\right],\, U_0=\frac{G M}{R}.
 \end{eqnarray}
 The governing equations for non-compressional irrotational vibrations of the homogeneous liquid star are
 \begin{eqnarray}
 \label{e4.8}
 &&\rho\delta {\dot v}_i=-
 \nabla_i\delta p+\rho\nabla_i\delta U+\nabla_k \delta\pi_{ik},\\
 \label{e4.9}
 &&\delta {\dot p}=-(\delta v_k\nabla_k)\,p,\quad \nabla^2 \delta U=0,\\
 \label{e4.10}
 && \delta \pi_{ik}=2\eta \delta v_{ik},\quad
 \delta v_{ik}=\frac{1}{2}(\nabla_i\delta v_k+\nabla_k\delta v_i),\quad \eta=\eta_c\left[1-\left(\frac{r}{R}\right)^2\right].
 \end{eqnarray}
 The equation of energy conservation is
 \begin{eqnarray}
 \label{e4.11}
 &&\frac{\partial }{\partial t}\int \frac{\rho\delta v^2}{2}\,d{\cal V}=
 -\int [(\delta v_k\nabla_k)\delta p - \rho (\delta v_k\nabla_k) \delta U
 -\delta\pi_{ik}\delta v_{ik}]\,
  d{\cal V}.
 \end{eqnarray}
 On substituting here
 $$\delta v_i({\bf r},t)=a_i({\bf r}){\dot \alpha} (t),\quad
 \delta U({\bf r},t)=\phi_i({\bf r}){\alpha}(t),\quad \delta p({\bf r},t)=-(a_k\nabla_k)\,p(r){\alpha}(t) $$
 we arrive at equation of damped oscillator, ${\cal M}{\ddot \alpha}+{\cal D}{\dot \alpha}+{\cal K}\alpha=0$, with integral parameters defining the frequency and lifetime
 of the form
  \begin{eqnarray}
 \label{e4.12}
 {\cal M}&=&\int_{\cal V}\rho\, a_k \,a_k\,d{\cal V},\quad
 {\cal D}=2\int \eta(r) a_{ik}\,a_{ik}\,d{\cal V},\\
 \label{e4.13}
 {\cal K}&=&-\int_{\cal V} [(a_i\nabla_i)\,(a_k\nabla_k) p(r)+\rho\,(a_k\,\nabla_k)\phi({\bf r})]\,d{\cal V},\\
 && \label{e4.14}
 \omega^2=\frac{{\cal K}}{{\cal M}},\quad
  \tau=\frac{2{\cal M}}{{\cal D}}.
 \end{eqnarray}
 The only unknown quantity is the variations of the gravity potentials obeying the
 Laplace equations
 \begin{eqnarray}
 \label{e4.15}
 \nabla^2 \delta U_i=0,\quad\quad \nabla^2 \delta U_e=0
 \end{eqnarray}
 having the general solutions of the form
 \begin{eqnarray}
 \label{e4.16}
 \delta U_i(r<R)=C_\ell\,r^\ell P_\ell(\zeta)\alpha(t), \quad\quad
 \delta U_e(r<R)=D_\ell\,r^{-(\ell+1)}P_\ell(\zeta)\alpha(t).
 \end{eqnarray}
 The arbitrary constants $C_\ell$ and $D_\ell$ are eliminated from
 the standard boundary conditions
 \begin{eqnarray}
 \label{e4.17}
 && U_i(r(t))+\delta U_i(r(t))=U_e(r(t))+\delta U_e(r(t))\Big\vert_{r=R},
 \\[0.2cm]
 \label{e4.18}
 && \frac{d}{dr}
 \left[U_i(r(t))+\delta U_i(r(t))\right]=
 \frac{d}{dr}\left[U_e(r(t))+\delta U_e(r(t))\right]\Big\vert_{r=R}
 \end{eqnarray}
 where $r(t)=r[1+ \alpha(t) P_\ell (\cos\theta)]$. Retaining in these equations terms of first order in $\alpha(t)$ and putting $r=R$ we arrive at coupled algebraic equations for $C_\ell$ and $D_\ell$ whose solution leads to the following final expressions for the time-independent part of gravity potential\cite{B-96b}
 \begin{eqnarray}
 \label{e4.19}
&&  \phi_i=\frac{4\pi}{2l+1}\frac{G\rho}{R^{\ell-2}}
 \,r^\ell\,P_\ell(\zeta),\quad
  \phi_e=\frac{4\pi}{2\ell+1}G\rho\,R^{\ell+3}
 \,r^{-(\ell+1)}P_\ell(\zeta).
 \end{eqnarray}
Computation of integral parameters of the inertia ${\cal M}$ viscous friction ${\cal D}$ yields
 \begin{eqnarray}
 \label{e4.20}
 && {\cal M}=4\pi\,A^2_\ell\,\rho R^{2\ell+1}\frac{\ell}{2\ell+1}=\frac{4\pi\rho R^5}{\ell\left(2\ell +1\right)},
 \\[0.2cm]
 \label{e4.21}
 && {\cal D}=16\pi\,A_\ell^2\eta_c R^{2\ell-1}\frac{\ell(\ell-1)}{2\ell+1}=16\pi \eta_c R^3\frac{(\ell-1)}{2\ell+1}.
 \end{eqnarray}
 One can consider separately oscillations restored by force associated
 with gradient in fluctuations of pressure and force owing its origin to fluctuations in the potential of
 gravity. In accord with this, the integral parameter of stiffness is written as a sum
\begin{eqnarray}
\label{e4.22}
&& {\cal K}={\cal K}_p+{\cal K}_g,\\
\label{e4.23}
&& {\cal K}_p=-\int_{\cal V}(a_k\nabla _k)\,(a_i\nabla _i)p(r)\,d{\cal V},\quad {\cal K}_g=-\int_{\cal V}
 \rho(a_i\nabla _i)\phi({\bf r})\,d{\cal V}
\end{eqnarray}
and, hence, for the squared frequency one has
\begin{eqnarray}
\label{e4.24}
&& \omega^2=\omega_p^2+\omega_g^2,\quad\omega_p^2=\frac{{\cal K}_p}{\cal M},\quad \omega_g^2=\frac{{\cal K}_g}{\cal M}
\end{eqnarray}
where $\omega_p$ designate the frequency of $p$-mode and $\omega_g$ is the frequency of $g$-mode.
For ${\cal K}_p$ and ${\cal K}_g$ we get
\begin{eqnarray}
\label{e4.25}
{\cal K}_p=\frac{16\pi^2}{3}\frac{G\rho^2R^5}{2\ell+1},\quad\quad {\cal K}_g=-\frac{16\pi^2 G\rho^2 R^5}{(2\ell+1)^2}.
\end{eqnarray}

\begin{figure}[h]
 \centering{ \includegraphics[width=8cm]{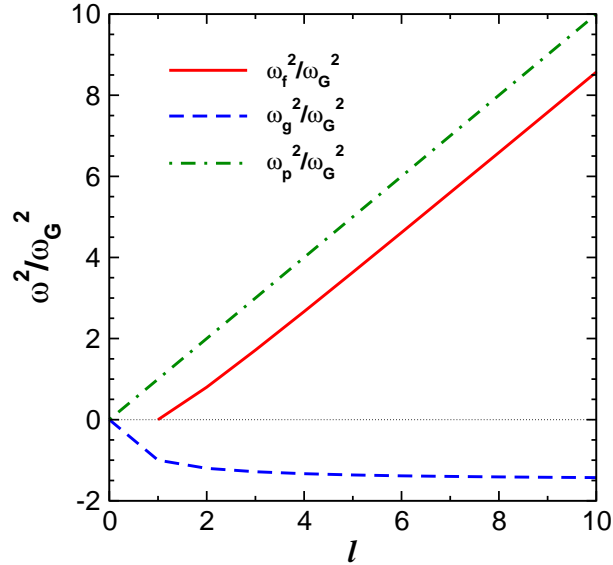}}
 \caption{
 The squared fractional frequencies of $p$-mode, $g$-mode and Kelvin's $f$-mode as functions of multipole degree
 $\ell$, computed in the homogeneous liquid star model.}
\end{figure}

For the frequencies of $p$-mode and $g$ mode we obtain
\begin{eqnarray}
\label{e4.26}
&& \omega_p^2=\frac{{\cal K}_p}{\cal M}=\omega_G^2\,\ell,\quad\quad
\omega_g^2=\frac{{\cal K}_g}{\cal M}=-\omega_G^2\frac{3\ell}{2\ell+1}\\
\label{e4.26A}
&& \omega_K^2(_0g^f_\ell)=\omega_p^2+\omega_g^2=\omega_G^2\,\frac{2\ell(\ell-1)}{(2\ell+1)}.
\end{eqnarray}
As is illustrated in Fig. 3, $\omega_p^2$ is positive, whereas $\omega_g^2$ is negative and
these sings for $p$-mode and $g$-mode have one and the same sense as in well-known dispersion relation of Jeans $\omega^2=c_s^2k^2-4\pi G\rho$ characterizing
propagation of longitudinal acoustic wave with the velocity of sound $c_s$ in a self-gravitating homogeneous fluid\cite{CH-61,T-04}.
 For the lifetime of $\ell$-pole overtone, $\tau_f(\ell)$, computed with the
 non-uniform radial profile of shear viscosity $\eta=\eta(r)$ given by equation (\ref{e4.10}), we obtain
\begin{eqnarray}
 \label{e4.27}
 && \tau_f(\ell)=\frac{\tau_\nu}{2(l-1)},\quad\quad
 \tau_\nu=\frac{R^2}{\nu},\quad \nu=\frac{\eta_c}{\rho}.
 \end{eqnarray}

 \begin{figure}[h]
 \centering{ \includegraphics[width=11cm]{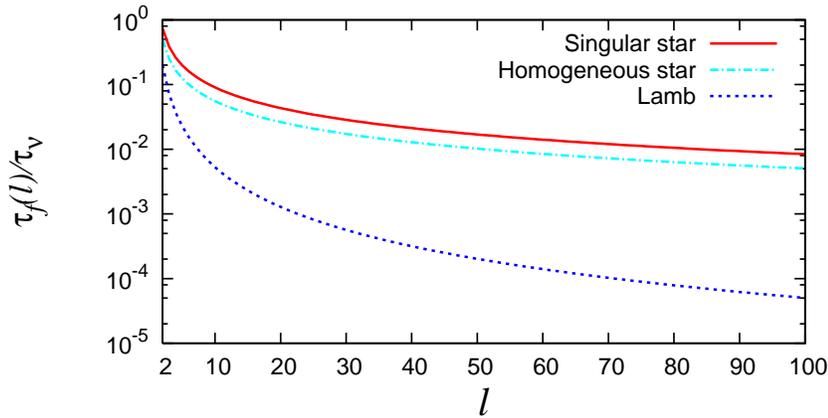}}
 \caption{
 The fractional lifetime of $\ell$-pole overtones of fundamental vibrational mode in homogeneous and inhomogeneous liquid star models with non-uniform profiles of shear viscosity in juxtaposition
 with Lamb spectral equation for damping time of nodeless irrotational oscillations of a spherical mass
 of viscous liquid of uniform density and constant coefficient of shear viscosity.}
\end{figure}

It is worthwhile to compare the computed lifetime spectra with the Lamb spectral equation\cite{LAMB}
\begin{eqnarray}
 \label{e4.28}
 && \tau_{\rm Lamb}(\ell)=\frac{\tau_\nu}{(2\ell+1)(l-1)}
 \end{eqnarray}
 that has been obtained in a similar fashion but assuming that coefficient of shear viscosity has
 constant value in the entire spherical volume of homogeneous viscous liquid\cite{IJMPA-07}.
 It has been pointed out by Jeffreys\cite{Jef-76}, however, in the context of geoseismology that
 the approximation of uniform viscosity
 does not allows for the effect of self-gravity on mechanical property of matter of astrophysical objects.
 With this in mind, we have supposed that it would be not inconsistent to take radial
 profile of shear viscosity
 similar to that for hydrostatic pressure in the star. Also noteworthy is that Newtonian law of shear viscosity is equally appropriate
 for viscous liquid and viscoelastic solid\cite{LL-7}. The  spectral
 equations for the time of viscose damping of nodeless spheroidal vibrations obtained hear for the first time may be of some interest,
 therefore, for the general seismology of Earth-like planet.
 In Fig.4, the obtained spectral equations for the lifetime, normalized to $\tau_\nu$, as a function of overtone number $\ell$ are plotted for both singular (\ref{e3.20}) and homogeneous (\ref{e4.27}) models
 in juxtaposition with the Lamb spectral formula (\ref{e4.28}).

\begin{figure}[h]
 \centering{\includegraphics[width=8cm]{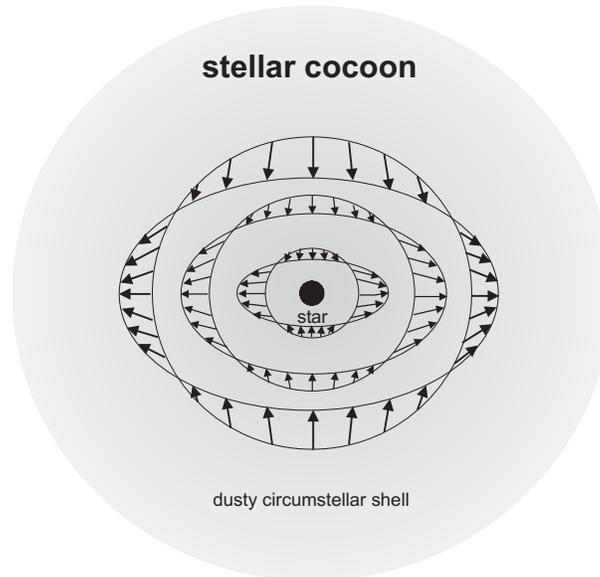}}
 \caption{Artist view of a spherical star-forming object, the stellar cocoon, whose
 largescale vibrations in $f$-mode can be analyzed on the basis considered highly inhomogeneous model.}
 \end{figure}

\section{Summary}

 The most striking differences between vibrational behavior of homogenous and inhomogeneous liquid star models in the fundamental mode of global nodeless irrotational pulsations
 under the combined action of the buoyancy and gravity forces is that in the inhomogeneous model the oscillations are of substantially compressional character, that is, accompanied by fluctuations in density and the lowest overtone of $f$-mode is of dipole degree, as has been first been observed by Smeyers (as pointed out in Ref.[8]) and in Ref.[16]. In the homogeneous star model
 they are characterized as non-compressional and the lowest overtone is of quadrupole degree.

 The obtained spectral equations for the frequency and life-time of fundamental mode
 in stellar object with a highly inhomogeneous density profile
 can be invoked in assessing  variability of emission from star-forming clouds like stellar cocoon, pictured in Fig.5, and too from red giants being on the pre-white dwarf stage as being produced by their global nodeless vibrations.
 These are stellar objects whose asteroseismology
 is adequately modeled by equations of fluid-dynamics, contrary to the extremely dense compact
 objects of finale stage of evolution track, white dwarfs and pulsars, the seismic
 vibrations of which are studied within the framework of solid star models relying on equations of
 solid mechanics, as  reported in Refs. [13,17-19]. Understandably, therefore, that mathematical treatment
 of node-free vibrations of a solid star in fundamental mode is different from the considered here
 fluid-dynamical theory of fundamental mode in liquid star models and this difference will be the subject our forthcoming paper.

\section*{Acknowledgments}

 The authors are grateful to Gwan-Ting Chen (NTHU, Taiwan) and Dima Podgainy (JINR, Russia) for helpful assistance. This work is a part of projects on investigation of variability of high-energy emission from compact sources
 supported by NSC of Taiwan, grant numbers  NSC-96-2628-M-007-012-MY3 and NSC-97-2811-M-007-003.

\appendix

\section{Appendices}

The integral parameter of stiffness
 \begin{eqnarray}
 \nonumber
 && {\cal K}=\int[(a_k\nabla_k\,\rho(r))(a_k\,\nabla_k\,U({\bf r}))-(a_k\nabla_k)\,(a_k\nabla_k) p(r)-\rho(r) (a_k\nabla_k)\phi({\bf r})]\,d{\cal V}
 \end{eqnarray}
can be conveniently represented in the form ${\cal K}=K_1+K_2+K_3$, where
\begin{eqnarray}
 \nonumber
  K_1&=&\int(a_k\,\nabla_k\,\rho(r))(a_k\,\nabla_k\,U({r}))d{\cal V}=
 \int\left(a_r(r,\theta)\frac{\partial \rho(r)}{\partial r}\right)
 \left(a_r(r,\theta)\frac{\partial U(r)}{\partial r}\right)d{\cal V}\\
 \nonumber
 &=&\frac{10\pi^2}{9}G\rho_s^2A_\ell^2\,R\,\ell^2
 \int\limits_{0}^{R} r^{2\ell-1}dr\,\int\limits_{-1}^{+1} P_\ell^2(\zeta)d\zeta=
 \frac{10\pi^2}{9}G\rho_s^2\,A_\ell^2\,R^{2\ell+1}\,\frac{\ell}{2\ell+1}.
 \end{eqnarray}
The integral for $K_2$ is given by
\begin{eqnarray}
\nonumber
 K_2&=&-\int(a_k\nabla_k)\,(a_k\nabla_k) p(r)\,d{\cal V}=-\int
 \left[a_r(r,\theta)\,\frac{\partial }{\partial r}+\frac{a_\theta(r,\theta)}{r}\frac{\partial }{\partial \theta} \right]\left(a_r(r,\theta)\,\frac{\partial p(r)}{\partial r}\right)\,d{\cal V} \\
 \nonumber
 &=&\frac{10\pi^2}{9}G\rho_s^2R\int\limits_{0}^{R}r^{2\ell-1}dr\,
  \left[\ell^2(\ell-1)\int\limits_{-1}^{+1}P_\ell^2(\zeta)d\zeta+
  \ell\int\limits_{-1}^{+1}(1-\zeta^2)^{1/2}\left(\frac{dP_\ell(\zeta)}{d\zeta}\right)^2d\zeta\right]\\
  \nonumber
  &=&\frac{40\pi^2}{9}A_\ell^2G\rho_s^2R^{2\ell+1}\frac{\ell^2}{2\ell+1}.
 \end{eqnarray}
In similar fashion, for $K_3$ we obtain
\begin{eqnarray}
  \nonumber
 K_3&=&-\int \rho(r) (a_k\,\nabla_k)\phi({\bf r})\,d{\cal V}=-\int \rho(r)
 \left[a_r(r,\theta)\,\frac{\partial \phi(r,\theta)}{\partial r}+\frac{a_\theta(r,\theta)}{r}\frac{\partial \phi(r,\theta)}{\partial \theta}\right]\,d{\cal V}\\
 \nonumber
 &=&\frac{-50\pi^2}{9} G \rho_0^2 A_{\ell}^2 R
\left[\int\limits_{0}^{R}\left(\frac{\ell ^3 r^{2\ell-5/2}}{\left(2\ell+1\right)R^{1/2}}-\ell ^2\frac{2\ell-1}{4\ell+1}r^{2\ell-3}\right)r^2 dr\int\limits_{-1}^1P_{\ell}^2(\zeta)\left(\zeta\right)d\zeta
\right. \\ \nonumber
 & & \qquad\qquad\qquad +  \left.
\int\limits_{0}^{R} \left(\frac{\ell r^{2\ell -5/2}}{\left(2\ell +1\right)R^{1/2}}-\frac{2\ell r^{2\ell-3}}{4\ell+1}\right)r^2 dr\int\limits_{-1}^1\left(1-\zeta^2\right)\left(\frac{dP_{\ell}(\zeta)}{d\zeta}\right)^2 d\zeta\right] \\
\nonumber
 &=&-\frac{50\pi^2}{9}A_\ell^2G\rho_s^2R^{2\ell+1}\frac{\ell}{(2\ell+1)(4\ell+1)}.
 \end{eqnarray}
The resultant expression for stiffness ${\cal K}$ reads
\begin{eqnarray}
\nonumber
{\cal K}=\frac{40\pi^2}{9}A_\ell^2G\rho_s^2R^{2\ell+1}\frac{\ell(4\ell^2+2\ell-1)}{(2\ell+1)(4\ell+1)}.
 \end{eqnarray}


\begin{thebibliography}{100}


\bibitem{K-1863} W. Thompson W (Kelvin), {\it Phil. Trans. Roy. Soc. Lond.} {\bf 153}, 384 (1863).

 \bibitem{LAMB} H. Lamb, {\it Hydrodynamics} (Dover, New York, 1945).

\bibitem{CH-61} S. Chandrasekharm {\it Hydrodynamic and Hydromagnetic Stability}
                            (Clarendon, Oxford, 1961).

\bibitem{Chandra-64} S. Chandrasekhar, {\it Astrophys. J.} {\bf 139},  664 (1964).

\bibitem{AS-77}  M. L. Azienman and P. Smeyers, {\it Astrophys and Space Sci.} {\bf 48},  123 (1970).


\bibitem{B-96a} S. I.  Bastrukov,   {\it Phys. Rev.} E {\bf 53}, 1917 (1996).

\bibitem{B-96b} S. I. Bastrukov, {\it Int. J. Mod. Phys.} D {\bf 5}, 45 (1996).

 \bibitem{COW-41}  A. Gautschy, {\it Vistas in Astronomy} {\bf 41}, 95 (1997).

\bibitem{CL-86}   D. D. Clayton, {\it Am. J. Phys.} {\bf 54}, 354 (1986).



\bibitem{HKT-04} C. J. Hansen, S. D. Kawaler and V. Trimble, {\it Stellar Interiors} 2nd edn. (Springer-Verlag, New York, 2004).

\bibitem{T-04} M. J. Thompson, {\it An Introduction to Astrophysical Fluid Mechanics} (Imperial College Press, World Scientific, 2004).

\bibitem{PHIL-94} A. C.  Phillips, {\it The Physics of Stars} (Wiley, New York, 1994).


\bibitem{IJMPA-07}  S. I. Bastrukov, H.-K. Chang, \c S. Mi\c sicu, I. V. Molodtsova and D. V. Podgainy,
                     {\it Int. J. Mod. Phys.} A {\bf 22},  3261 (2007).

\bibitem{Jef-76}  H. Jeffreys, {\it The Earth} 6th ed. (Cambridge University Press, 1976).


\bibitem{LL-7}  L. D. Landau, E. M. Lifshits, A. M. Kosevich and L. P. Pitaevskii, {\it Theory of
                  Elasticity} 3d edn. (Pergamon, Oxford,  1986).



\bibitem{PBAST-96} D. V. Podgainy, S. I. Bastrukov,  I. V. Molodtsova and V. V. Papoyan,
                    {\it Astrophys.} {\bf 39},  278 (1996).


\bibitem{MNRAS-07} S. I. Bastrukov, H.-K. Chang, J. Takata, G.-T. Chen and I. V. Molodtsova, {\it Mon. Not.
 R. Astron. Soc.} {\bf 382},  849 (2007).

\bibitem{MPL-08} S. I. Bastrukov, H.-K. Chang, G.-T. Chen, and I. V. Molodtsova, {\it Mod. Phys. Lett.} A {\bf 23}, 477 (2008).

\bibitem{ApJ-08} S. I. Bastrukov, G.-T. Chen, H.-K. Chang, I. V. Molodtsova and D. V. Podgainy,
 {\it Astrophys. J.}, 690,  998 (2009).



\end{thebibliography}
\end{document}